\documentclass[referee]{aa}

\usepackage{graphicx}
\usepackage[varg]{txfonts}
\usepackage{bm}
\usepackage{booktabs}
\usepackage{hyperref}
\hypersetup{hidelinks}
\begin{document}

\title{Strong-lensing tests of angular structure in the radial acceleration relation}
\titlerunning{Strong-lensing tests of angular RAR structure}
\authorrunning{C. Alard}

\author{C. Alard\inst{1}}
\institute{Institut d'Astrophysique de Paris, 98bis, boulevard Arago, 75014 Paris}
\date{}

\abstract
{The radial acceleration relation (RAR) links the radial acceleration inferred from galaxy rotation curves to that predicted from the observed baryonic distribution. Whether it has a counterpart in the angular structure of gravitational fields remains unknown.}
{We develop an observational framework to test for such a non radial correspondence using strong gravitational lenses, without assuming that a generalized RAR exists.}
{The singular perturbative approach reconstructs two angular fields from lensed image morphology without imposing a parametric mass profile. We separate contributions from projected mass inside and outside the Einstein circle, normalize the inner baryonic and total fields, and define statistical comparisons of their radial and tangential components. We also derive higher order potential multipoles for a smooth elliptical, off centered projected halo and evaluate their dependence on NFW and Einasto profiles.}
{The inner outer separation isolates the projected internal contribution most directly relevant to the RAR comparison. The inner dipole coefficients can be recovered without the source position contribution when both reconstructed fields are used. The quadrupole phase provides a measure of projected orientation, while the higher order moments quantify departures from a smooth elliptical profile. For the profiles and parameter ranges considered, the predicted higher order ratios remain small and vary only moderately with the Einstein radius to scale radius ratio.}
{The framework provides a model independent observational test of whether a three dimensional non radial baryonic Total field correspondence leaves a statistical imprint after projection. The large strong lens population expected from Euclid could provide the sample needed for this test, subject to image quality, photometric, and selection function requirements.}

\keywords{gravitational lensing: strong -- galaxies: kinematics and dynamics -- galaxies: structure -- dark matter -- methods: analytical -- cosmology: observations}

\maketitle

\section{Introduction}
\label{sec:intro}

The radial acceleration relation (RAR) connects the acceleration inferred from galaxy rotation curves to that predicted from the observed baryonic distribution \citep{McGaugh2016}. Its small scatter over a broad range of galaxy properties makes it an important empirical constraint on galactic dynamics. Its physical origin remains unsettled, independently of any particular theoretical interpretation.

The RAR is formulated in terms of radial accelerations. A galaxy's gravitational field, however, is not generally radial: the baryonic distribution is not perfectly asymmetric, and the additional gravitating component need not share its shape or orientation. It is therefore natural to ask whether the RAR has a counterpart in the angular structure of the gravitational field. Such a relation, if present, could provide information not contained in radial profiles alone. A null result, or a dependence on morphology and environment, would also constrain the physical meaning and domain of validity of the RAR.

Strong gravitational lensing provides access to the angular structure of the projected gravitational potential near the Einstein radius. Extended arcs and multiple images respond to departures from circular symmetry. The singular perturbative approach developed by \citet{Alard2007,Alard2008,Alard2009,Alard2010} is particularly suited to this problem: it reconstructs two angular fields directly connected to the observed image morphology, without imposing a parametric density profile or a prescribed angular form for the lens.

The present work develops this framework as an observational test, not as a derivation of a non radial relation from a chosen theory. We first clarify what lensing measures and the acceleration range probed by existing galaxy scale lenses. We then introduce the singular perturbative description using the notation of the original papers, and discuss the mass sheet degeneracy and the normalization required for comparisons between total and baryonic fields. The multipole expansion is introduced only after these fields and their normalization have been defined. We then separate inner and outer contributions, construct two statistical RAR like diagrams from the normalized inner fields, and examine the quadrupole and higher order multipoles. The Discussion closes with the statistical opportunity offered by Euclid.

\section{The projected gravitational field}
\label{sec:projection}

Gravitational lensing does not measure the local three dimensional gravitational acceleration at a point in the lens. It probes derivatives of the gravitational potential projected along the line of sight. Let $\Phi(x,y,z)$ be the three dimensional Newtonian potential, with the $z$ direction parallel to the line of sight. The transverse derivatives relevant to lensing are defined by
\begin{equation}
\frac{\partial\Psi}{\partial x}\equiv
\int_{-\infty}^{+\infty}\frac{\partial\Phi}{\partial x}\,dz,
\qquad
\frac{\partial\Psi}{\partial y}\equiv
\int_{-\infty}^{+\infty}\frac{\partial\Phi}{\partial y}\,dz.
\label{eq:projected_gradient}
\end{equation}
The notation $\Psi$ is used for the projected potential derivatives; the liefest integral of $\Phi$ itself need not converge. Since $\bm{g}=-\bm{\nabla}\Phi$, the projected transverse gradient is the line of sight integral of the transverse gravitational field,
\begin{equation}
\bm{\nabla}_{\perp}\Psi=-\int_{-\infty}^{+\infty}\bm{g}_{\perp}\,dz.
\label{eq:projected_field}
\end{equation}
This is a reduced two dimensional projected field. The line of sight integral of an acceleration has dimensions of velocity squared, so it must not be identified with the local three dimensional acceleration entering the usual RAR. In the following, ``reduced field'' refers to the two dimensional quantities reconstructed in the lens plane, with this projection understood.

The convergence of the projected derivatives is sufficient for the lensing calculation; convergence of the line of sight integral of $\Phi$ itself is not required. For a finite, sufficiently regular halo, the first and second derivatives of $\Phi$ have sufficiently rapid asymptotic decay for their line of sight integrals to converge. The proof is given in Appendix~\ref{app:convergence}. Integrating the three-dimensional Poisson equation along $z$ then gives
\begin{equation}
\frac{\partial^2\Psi}{\partial x^2}
+\frac{\partial^2\Psi}{\partial y^2}
=4\pi G\,\Sigma(x,y),
\qquad
\Sigma(x,y)=\int_{-\infty}^{+\infty}\rho(x,y,z)\,dz,
\label{eq:projected_poisson}
\end{equation}
where the boundary term $[\partial_z\Phi]_{-\infty}^{+\infty}$ vanishes. The reconstructed lensing field is thus directly related to the 2D projection of the 3D accelerations.

\section{Characteristic acceleration range of galaxy scale lenses}
\label{sec:range}

The Einstein radii of galaxy scale strong lenses typically lie in the inner regions of massive galaxies, where the baryonic and dark components can both contribute significantly to the gravitational field. To indicate the characteristic acceleration regime sampled by existing systems, we estimate the acceleration at the Einstein radius using a singular isothermal sphere (SIS) approximation:
\begin{equation}
a_{\rm SIS}(R_E)=\frac{2\sigma^2}{R_E},
\qquad R_E=D_l\theta_E,
\label{eq:sis_acceleration}
\end{equation}
where $\sigma$ is the velocity dispersion, $D_l$ is the angular diameter distance to the lens, and $\theta_E$ is its angular Einstein radius.

For the 25 grade-A systems in the SL2S sample for which the required lensing and spectroscopic parameters are available, this estimate gives approximately
\begin{equation}
1.1\lesssim \frac{a_{\rm SIS}(R_E)}{a_0}\lesssim 10,
\qquad {\rm median}\simeq 4.6,
\end{equation}
where we adopt $a_0=1.2\times10^{-10}\,\mathrm{m\,s^{-2}}$ and a flat $\Lambda$CDM cosmology with $H_0=70\,\mathrm{km\,s^{-1}\,Mpc^{-1}}$ and $\Omega_m=0.3$. The lensing and photometric quantities are taken from \citet{Sonnenfeld2013a}, while the spectroscopic measurements are presented by \citet{Sonnenfeld2013b}. The 25 systems constitute the subset for which the necessary redshifts, Einstein radii, and velocity dispersions are jointly available.

These values are SIS-equivalent estimates intended to characterize the acceleration range, not precise measurements of the local three-dimensional acceleration. They do not represent the reduced projected fields reconstructed below. Their purpose is to show that galaxy scale strong lenses probe characteristic accelerations of interest for the RAR, including systems relatively close to its empirical acceleration scale. Existing samples therefore provide a useful basis for developing the method, while a much larger and more homogeneous sample is needed for population statistics.

\section{The singular perturbative reconstruction}
\label{sec:method}

We use the singular perturbative approach developed by \citet{Alard2007} and applied to observed lenses in \citet{Alard2008,Alard2009,Alard2010}. The method expands the lens around a circularly symmetric critical configuration. In the dimensionless lens plane coordinates used below, the potential is written
\begin{equation}
\phi(r,\theta)=\phi_0(r)+\psi(r,\theta),
\label{eq:phi_decomposition}
\end{equation}
where $\phi_0$ is the circular reference potential and $\psi$ is a small non anti-symmetric perturbation. We set the Einstein radius to $R_E=1$ and normalize the reference radial field so that $\phi_0'(1)=1$.

The perturbative fields are defined on the Einstein circle by
\begin{equation}
\left.\frac{\partial\psi}{\partial r}\right|_{r=1}=f_1(\theta),
\qquad
\left.\frac{\partial\psi}{\partial\theta}\right|_{r=1}=f_0'(\theta),
\label{eq:alard_fields}
\end{equation}
where $f_0'=df_0/d\theta$. Thus the gradient of the lensing potential is
\begin{equation}
\left.\bm\nabla\phi\right|_{r=1}
=[1+f_1(\theta)]\bm e_r+f_0'(\theta)\bm e_\theta.
\label{eq:full_gradient}
\end{equation}
We denote the perturbative departure from the circular reference field by
\begin{equation}
\bm F(\theta)=f_1(\theta)\bm e_r+f_0'(\theta)\bm e_\theta.
\label{eq:perturbative_vector}
\end{equation}
The unit radial term in Eq.~(\ref{eq:full_gradient}) is the reference circular field; it is not part of $\bm F$.

The fields $f_0$ and $f_1$ are directly connected to the morphology of the lensed images. Their reconstruction follows from the observed arcs and image configuration; it does not require a parametric mass profile, a prescribed halo shape, or an assumed external shear form. The practical reconstruction of the lens and source from observed arcs, including an initial solution and its subsequent refinement, is described by \citet{Alard2024}. The choice of the minimal Fourier order is also important: unnecessary higher order terms can be generated by source lens coupling and need not represent physical structure \citep{Alard2023}. The detailed algorithms are given in these papers and are not repeated here. In this sense the method reconstructs the angular fields themselves rather than selecting a mass model from a family of parameterized profiles. The mass sheet degeneracy nevertheless remains, and we discuss below how it can be addressed.

\section{Mass sheet degeneracy and normalization}
\label{sec:degeneracy}

\subsection{The perturbative fields}
\label{sec:field_normalization}

The mass sheet degeneracy has a simple manifestation in the singular perturbative formalism. As discussed by \citet{Alard2007}, it re-scales the two perturbative fields $f_0$ and $f_1$ by a common factor and produces the corresponding rescaling of the source. We use this established result without re deriving it.

Since the vector $\bm F$ in Eq.~(\ref{eq:perturbative_vector}) is multiplied by the same factor, its normalized angular structure is invariant. We define
\begin{equation}
\|\bm F\|^2=\frac{1}{2\pi}\int_0^{2\pi}
\left[f_1^2(\theta)+f_0'^2(\theta)\right]d\theta,
\qquad
\widehat{\bm F}(\theta)=\frac{\bm F(\theta)}{\|\bm F\|}.
\label{eq:field_normalization}
\end{equation}
The normalization removes the common amplitude while retaining the angular pattern. It is this pattern, rather than the absolute perturbative amplitude, that is used in the comparisons below. The same normalization can be applied after separating the inner and outer contributions, since that decomposition is linear in the perturbative fields.

\subsection{The baryonic field}
\label{sec:baryonic_normalization}

The baryonic field is obtained from the observed light distribution, projected into the lens plane and treated with the same Einstein radius normalization. If $\bm F_L$ denotes the perturbative field calculated for unit stellar mass to light ratio, a constant global mass to light ratio $\Upsilon$ gives
\begin{equation}
\bm F_{\rm bar}(\theta)=\Upsilon\bm F_L(\theta).
\end{equation}
Its normalized form is therefore
\begin{equation}
\widehat{\bm F}_{\rm bar}(\theta)
=\frac{\bm F_L(\theta)}{\|\bm F_L\|},
\label{eq:baryonic_normalization}
\end{equation}
which is independent of the global value of $\Upsilon$. This does not remove changes in the field shape caused by spatial variations in the stellar mass to light ratio, dust, or uncertainty in the baryonic components. The normalization is applied separately to the total and baryonic fields; it therefore tests their angular structures, not their relative amplitudes.

\section{Multipole expansion and inner outer separation}
\label{sec:potential_multipoles}

We now introduce the multipole expansion of $\psi$ and define the coefficients used below. The axisymmetric part of the projected potential is included in the circular reference potential. The non-axisymmetric perturbing potential can be separated into contributions generated by projected mass inside and outside the unit Einstein circle. In the usual dimensionless lensing normalization, its multipole expansion is
\begin{align}
\psi_{\rm in}(r,\theta)
&=-\sum_{n=1}^{\infty}\frac{a_n\cos(n\theta)+b_n\sin(n\theta)}{r^n},
\qquad r\geq1,\label{eq:psi_inner_multipoles}\\
\psi_{\rm out}(r,\theta)
&=-\sum_{n=1}^{\infty}r^n
\left[c_n\cos(n\theta)+d_n\sin(n\theta)\right],
\qquad r\leq1.
\label{eq:psi_outer_multipoles}
\end{align}
The first expression is the harmonic potential generated by matter inside the Einstein circle, evaluated outside that matter distribution; the second is the harmonic potential generated by matter outside the circle, evaluated inside that distribution. Both expressions are to be evaluated at their common boundary $r=1$ when constructing the perturbative fields. The radial coordinate is measured in units of $R_E$.

The real coefficients are related to the dimensionless projected surface density (convergence) $\Sigma$ by
\begin{align}
a_n&=\frac{1}{2\pi n}\int_0^1 u^{n+1}\,du
\int_0^{2\pi}\Sigma(u,\theta)\cos(n\theta)\,d\theta,
\label{eq:an_density}\\
b_n&=\frac{1}{2\pi n}\int_0^1 u^{n+1}\,du
\int_0^{2\pi}\Sigma(u,\theta)\sin(n\theta)\,d\theta,
\label{eq:bn_density}\\
c_n&=\frac{1}{2\pi n}\int_1^{\infty} u^{1-n}\,du
\int_0^{2\pi}\Sigma(u,\theta)\cos(n\theta)\,d\theta,
\label{eq:cn_density}\\
d_n&=\frac{1}{2\pi n}\int_1^{\infty} u^{1-n}\,du
\int_0^{2\pi}\Sigma(u,\theta)\sin(n\theta)\,d\theta.
\label{eq:dn_density}
\end{align}
These are the standard multipole coefficients of the two dimensional lensing potential \citep{Kochanek1991a,Alard2008,Alard2009,Alard2010}. In particular, $a_n$ and $b_n$ are the inner coefficients, while $c_n$ and $d_n$ are the outer coefficients. The normalization by $2\pi n$ is part of this potential multipole convention and will be retained in all moment ratios below; we follow the convention used by \citet{Kochanek1991a} and in our earlier application of the perturbative method.

For compactness, we combine the two real inner coefficients into the complex moment
\begin{equation}
Q_n\equiv a_n+i b_n
=\frac{1}{2\pi n}\int_0^1u^{n+1}\,du
\int_0^{2\pi}\Sigma(u,\theta)e^{+in\theta}\,d\theta.
\label{eq:Qn_definition}
\end{equation}
Thus $a_n=\operatorname{Re}Q_n$ and $b_n=\operatorname{Im}Q_n$. For a particular matter component $X$ (for example, the baryonic or total projected distribution), the same definitions are applied to its surface density, giving $Q_{n,X,\mathrm{in}}=a_{n,X,\mathrm{in}}+i b_{n,X,\mathrm{in}}$. The corresponding complex combination of outer coefficients could be defined as $c_n+i d_n$, but we will keep the outer coefficients separate since they are strongly influenced by external shear.
\subsection{The reconstructed fields: inner and outer terms}
\label{sec:innerouter}

The inner coefficients $a_n,b_n$ describe projected mass inside the Einstein circle; $c_n,d_n$ describe projected mass outside it. This is a split by projected radius in the lens plane, not by three dimensional distance along the line of sight.

The relation between these coefficients and the two perturbative fields is
\begin{align}
f_1(\theta)&=\sum_{n\geq1}n\left[(a_n-c_n)\cos n\theta
+(b_n-d_n)\sin n\theta\right],
\label{eq:inner_outer_f1}\\
f_0'(\theta)&=\sum_{n\geq1}n\left[-(b_n+d_n)\cos n\theta
+(a_n+c_n)\sin n\theta\right].
\label{eq:inner_outer_f0}
\end{align}

The first harmonics require a brief comment because the source position also
contributes to them in the singular perturbative reconstruction
\citep{Alard2007}. Writing the measured first harmonics, including these
source position terms, as
\begin{equation}
\widetilde f_1^{(1)}(\theta)=A_c\cos\theta+A_s\sin\theta,
\qquad
\widetilde f_0'{}^{(1)}(\theta)=B_c\cos\theta+B_s\sin\theta,
\label{eq:first_harmonics_measured}
\end{equation}
the relations are
\begin{align}
A_c&=a_1-c_1+x_0, & A_s&=b_1-d_1+y_0,\\
B_c&=-(b_1+d_1)+y_0, & B_s&=a_1+c_1-x_0,
\label{eq:first_harmonics_source}
\end{align}
where $(x_0,y_0)$ denotes the source position contribution in the adopted
dimensionless coordinates. Combining the radial and tangential harmonics
therefore gives
\begin{align}
a_1&=\frac{A_c+B_s}{2},\qquad
b_1=\frac{A_s-B_c}{2},
\label{eq:inner_dipole_independent_source}\\
c_1&=\frac{B_s-A_c}{2}+x_0,\qquad
d_1=y_0-\frac{A_s+B_c}{2}.
\end{align}
The source position terms cancel exactly in the inner dipole coefficients,
but not in the outer ones. Thus, within this decomposition, the inner
dipole can be recovered without knowing the source position by using both
reconstructed fields, rather than either field alone.

It is therefore natural to define the inner and outer contributions to the reduced perturbative field separately:
\begin{align}
\bm F_{\rm in}(\theta)=\sum_{n\geq1}n\bigg\{&
[a_n\cos(n\theta)+b_n\sin(n\theta)]\bm e_r\nonumber\\
&+[-b_n\cos(n\theta)+a_n\sin(n\theta)]\bm e_\theta\bigg\},
\label{eq:Fin}
\\
\bm F_{\rm out}(\theta)=\sum_{n\geq1}n\bigg\{&
-[c_n\cos(n\theta)+d_n\sin(n\theta)]\bm e_r\nonumber\\
&+[-d_n\cos(n\theta)+c_n\sin(n\theta)]\bm e_\theta\bigg\}.
\label{eq:Fout}
\end{align}
With these definitions,
\begin{equation}
\bm F_{\rm tot}(\theta)=\bm F_{\rm in}(\theta)+\bm F_{\rm out}(\theta).
\label{eq:Fsum}
\end{equation}

\subsection{Why the inner multipoles are used for the RAR comparison}

The RAR concerns the internal gravitational field of a galaxy. In spherical symmetry, the radial acceleration at a given radius is determined by the mass enclosed within that radius. The inner multipoles are therefore the natural lensing quantities for comparison with the RAR: they describe the projected mass distribution inside the Einstein aperture. Their separation from the outer contribution was demonstrated for SL2S02176-0513 by \citet{Alard2010}. The correspondence is not exact, since lensing integrates along the line of sight and the aperture is a projected circle rather than a three-dimensional sphere. Among the contributions reconstructed by the perturbative method, however, the inner terms have the closest connection to the internal field relevant to the RAR.

The outer multipoles have a different role. Their leading quadrupole contains the tidal perturbation usually described as external shear; higher orders encode more complex projected structure from nearby galaxies, groups, clusters, and line of sight matter. The geometry of this outer mass distribution is often complex and uncertain, and the outer terms should not be mixed with the inner multipoles in the primary RAR comparison.

There is also a population selection effect. For common moderate shear configurations, external shear can enlarge the tangential caustic and increase the source plane cross section for quad production. A sample selected through strong lens detections can consequently be weighted toward systems with stronger external perturbations. This effect has been studied for singular isothermal lenses with external shear by \citet{LeeKim2014}. Its magnitude in a survey depends on the source population, magnification bias, image detection thresholds, and selection criteria. It is therefore a potential systematic bias in the observed outer multipole distribution, not simply an uncertainty in individual measurements.

\section{Normalized inner fields and RAR like diagrams}
\label{sec:normalization}

We now apply the normalization of Section~\ref{sec:degeneracy} to the inner contribution isolated in Section~\ref{sec:potential_multipoles}. For each lens, let
\begin{equation}
\begin{aligned}
\bm F_{X,\rm in}(\theta)&=F_{X,\rm in,r}(\theta)\bm e_r
+F_{X,\rm in,\theta}(\theta)\bm e_\theta,
\qquad X\in\{\mathrm{tot},\mathrm{bar}\},\\
\mathrm{tot}&:\ \text{total field},\qquad
\mathrm{bar}:\ \text{baryonic field}.
\end{aligned}
\label{eq:FXinner}
\end{equation}
where the total field is reconstructed from lensing and the baryonic field is calculated from the projected light distribution. The same inner/outer split is applied to both components. We normalize each inner field separately:
\begin{equation}
\|\bm F_{X,\rm in}\|^2=\frac{1}{2\pi}\int_0^{2\pi}
\left[F_{X,\rm in,r}^2(\theta)+F_{X,\rm in,\theta}^2(\theta)\right]d\theta,
\qquad
\widehat{\bm F}_{X,\rm in}=\frac{\bm F_{X,\rm in}}{\|\bm F_{X,\rm in}\|}.
\label{eq:normalized_inner_field}
\end{equation}

An ensemble of lenses can then populate two RAR like diagrams, one for each component of the reduced two dimensional field. At every angular position in every lens, the radial components define
\begin{equation}
\mathcal P_{r,\rm in}:\qquad
x=\widehat F_{\rm bar,in,r}(\theta),\qquad
y=\widehat F_{\rm tot,in,r}(\theta),
\label{eq:radial_map}
\end{equation}
and the tangential components define
\begin{equation}
\mathcal P_{\theta,\rm in}:\qquad
x=\widehat F_{\rm bar,in,\theta}(\theta),\qquad
y=\widehat F_{\rm tot,in,\theta}(\theta).
\label{eq:tangential_map}
\end{equation}
The signed components must be retained because their signs carry angular information. These diagrams compare angular patterns, not absolute accelerations: the fields are reduced two dimensional projected perturbations and are not the local three dimensional accelerations used in the original RAR.

The radial acceleration relation (RAR) establishes an empirical correspondence between the magnitudes of the total and baryonic radial accelerations. Extending this question beyond radial symmetry requires considering not only the amplitudes of the gravitational fields but also their angular structure. The normalized multipoles considered here provide a way to investigate this correspondence. In particular, the phase of the quadrupole characterizes the relative orientation of the projected total and baryonic acceleration fields, while higher order moments provide a more general description of their angular correspondence. The normalization deliberately removes information about the relative amplitudes of the fields, and therefore does not test the RAR as a functional relation between acceleration magnitudes. Instead, it isolates a complementary aspect of the problem: whether the angular structure of the total gravitational field exhibits systematic correspondence with that of the baryonic field. This provides an observational framework for investigating possible non radial regularities without assuming in advance that a generalized RAR exists.

Different lenses probe different Einstein radii and characteristic acceleration regimes. The diagrams can therefore be examined in bins of an independently estimated quantity such as $a_{\rm SIS}(R_E)/a_0$, or as a continuous function of that quantity. The SIS estimate is used only to label the radial regime; it is not identified with either coordinate in Eqs.~(\ref{eq:radial_map}) and (\ref{eq:tangential_map}). No functional form for a non-radial relation is imposed in advance.

Angular measurements within one lens are correlated and must not be counted as independent lens systems. A population analysis should treat each lens as the basic statistical unit or explicitly include within lens covariance. Sample selection must also be modeled, even though the inner/outer split removes the direct contribution of external shear from the primary fields.

\section{Information in the inner multipoles}
\label{sec:multipoles}

The normalized inner fields can be characterized by their multipole moments. The quadrupole describes the leading departure from circular symmetry; higher moments probe departures from a simple elliptical profile. We consider these in turn.

\subsection{Quadrupole normalization and orientation}
\label{sec:quadrupole}

In this section we consider the inner potential multipoles. For each projected mass component $X\in\{\mathrm{tot},\mathrm{bar}\}$, write
\begin{equation}
Q_{n,X,\rm in}=a_{n,X,\rm in}+i b_{n,X,\rm in}.
\end{equation}
We normalize the complete set of inner multipole coefficients of each component by the norm of its quadrupole,
\begin{equation}
\mathcal N_{2,X}=\left(a_{2,X,\rm in}^2+b_{2,X,\rm in}^2\right)^{1/2}
=|Q_{2,X,\rm in}|,
\label{eq:quadrupole_norm}
\end{equation}
so that
\begin{equation}
\widehat a_{n,X,\rm in}=\frac{a_{n,X,\rm in}}{\mathcal N_{2,X}},\qquad
\widehat b_{n,X,\rm in}=\frac{b_{n,X,\rm in}}{\mathcal N_{2,X}},\qquad
\widehat Q_{n,X,\rm in}=\frac{Q_{n,X,\rm in}}{\mathcal N_{2,X}}.
\label{eq:normalized_multipoles}
\end{equation}
In particular, $|\widehat Q_{2,X,\rm in}|=1$. The normalization removes the overall amplitude of the multipole sequence while preserving its angular structure. It is applied separately to the total and baryonic components. This is consistent with the normalization of the perturbative fields in Sect.~\ref{sec:degeneracy}: a common re-scaling of the inner coefficients, whether associated with the mass sheet degeneracy or with a constant baryonic mass to light ratio, cancels in Eq.~(\ref{eq:normalized_multipoles}). Ratios between moments of the same component are unchanged, for example
\begin{equation}
\frac{\widehat Q_{3,X,\rm in}}{\widehat Q_{1,X,\rm in}}
=\frac{Q_{3,X,\rm in}}{Q_{1,X,\rm in}},\qquad
\frac{\widehat Q_{4,X,\rm in}}{\widehat Q_{2,X,\rm in}}
=\frac{Q_{4,X,\rm in}}{Q_{2,X,\rm in}}.
\end{equation}
The normalization requires a non zero, sufficiently well measured quadrupole; systems with a very weak quadrupole must be treated separately because their normalized higher moments would be unstable.

The quadrupole phase gives the principal orientation of the corresponding projected distribution. Since
\begin{equation}
a_{2,X,\rm in}\cos2\theta+b_{2,X,\rm in}\sin2\theta
=|Q_{2,X,\rm in}|\cos\left(2\theta-\arg Q_{2,X,\rm in}\right),
\end{equation}
its position angle is
\begin{equation}
\varphi_{2,X,\rm in}=\frac12\arg\widehat Q_{2,X,\rm in}
=\frac12\arg Q_{2,X,\rm in},
\end{equation}
with the usual $\pi$ ambiguity of an ellipse. The relative orientation of the total and baryonic quadrupoles is
\begin{equation}
\Delta\varphi_{2,\rm in}
=\frac12\arg\left(
\widehat Q_{2,\rm tot,\rm in}\widehat Q_{2,\rm bar,\rm in}^{*}
\right),
\label{eq:delta_phi2}
\end{equation}
reduced to a chosen interval modulo $\pi$. The coefficient convention must be identical for the baryonic and total reconstructions.

The practical measurability of this orientation is illustrated by the analysis of SL2S02176-0513 by \citet{Alard2010}. The measured difference in inclination between the baryonic and total distributions was $1.5^\circ\pm5.5^\circ$, showing that a relative orientation can be constrained to an accuracy of about five degrees with high resolution HST imaging. This result provides an observational precedent for the measurement proposed here.
This angle is already an interesting observable. In the standard CDM picture, stellar and dark matter distributions need not have exactly the same shape or principal axes. Three dimensional hydrodynamical simulations find non zero misalignment's, whose distributions depend on halo mass, aperture, galaxy type, and the components being compared. For example, the mean stellar dark matter misalignment reported for the MassiveBlack-II simulation varies from about $30^\circ$ at sub halo masses near $10^{10}\,h^{-1}M_\odot$ to about $10^\circ$ near $10^{14}\,h^{-1}M_\odot$ \citep{Tenneti2014}. Analyses of EAGLE and cosmo-OWLS likewise show that the inferred alignment depends strongly on the radial aperture and on whether stars are compared with the whole halo or with the local total matter distribution \citep{Velliscig2015}. The Illustris analysis also illustrates the dependence of the stellar matter alignment on the galaxy sample and the scale considered \citep{Xu2017}.

These quoted angles refer to three dimensional shape or axis comparisons. They are included only to indicate plausible orders of magnitude for intrinsic misalignment; they are not predictions for the projected two dimensional angle $\Delta\varphi_{2,\rm in}$ measured here. Projection, aperture, halo mass, redshifts, and galaxy population all affect the relation between the intrinsic and projected angles. Moreover, the total inner quadrupole is the vector sum of the baryonic and dark matter inner quadrupoles,
\begin{equation}
Q_{2,\rm tot,\rm in}
=Q_{2,\rm bar,\rm in}
+Q_{2,\rm dm,\rm in},
\end{equation}
when the coefficients refer to the same linear projected density multipole convention. Its angle relative to the baryonic quadrupole is therefore not, in general, identical to the angle between the dark matter and baryonic quadrupoles; it also depends on their relative amplitudes. A statistical measurement of $\Delta\varphi_{2,\rm in}$ nevertheless provides a direct test of the alignment of projected internal structures and of how any misalignment varies across the lens population. The quadrupole orientation becomes poorly defined when its amplitude is very small; such systems require an amplitude dependent orientation uncertainty or separate treatment.

\subsection{Higher order moments of a smooth elliptical, off centered halo}
\label{sec:elliptical_moments}

We estimate the higher order moments expected from a smooth, mildly elliptical halo whose center is displaced relative to the adopted lens center. For simplicity, the model is constructed by deforming the projected two dimensional circular surface density profile; it is not obtained by deforming a three dimensional triaxial halo and then projecting it along the line of sight. This simplified prescription is intended only to estimate the order of magnitude of the multipoles generated by a smooth, mildly deformed projected halo, not to provide a realistic model of halo triaxiality. The quantity to be expanded is the inner potential multipole defined in Eq.~(\ref{eq:Qn_definition}). For the model considered here it is a function of the ellipticity parameter $\eta$ and the two coordinates of the displacement $(x_c, y_c)$, or $(r_0, \theta_0)$ in polar coordinates:
The projected density $\Sigma$ is thus expressed as:
\begin{equation}
\Sigma(x,y)=K\left((1-\eta) (x-x_c)^2+(1+\eta) (y-y_c)^2\right)=K\left(s(r,\theta)\right)
\label{eq:sigma}
\end{equation}
The inner potential multipole inside the Einstein radius ($R_E=1$) is then:
\begin{equation}
Q_n=\frac{1}{2\pi n}\int_0^1 r^{n+1}\,dr
\int_0^{2\pi}K\!\left(s(r,\theta)\right)e^{+in\theta}\,d\theta.
\label{eq:Qn_integral}
\end{equation}
 We take the ellipticity and the displacement to be of the same perturbative order, and introduce fixed bookkeeping amplitudes $\eta_0$, $x_{c,0}$, and $y_{c,0}$:
\begin{equation}
\eta_{\rm phys}=\epsilon\eta_0,\qquad
x_{c,\rm phys}=\epsilon x_{c,0},\qquad
 y_{c,\rm phys}=\epsilon y_{c,0}.
\label{eq:epsilon_parameters}
\end{equation}
For compactness, the subscript ``phys'' is suppressed after the expansion; $\eta$, $x_c$, and $y_c$ in the final moment formulae denote the physical ellipticity and displacement.
The inner multipole is consequently a function of the bookkeeping parameter $\epsilon$:
\begin{equation}
Q_n(\epsilon)=\frac{1}{2\pi n}\int_0^1 r^{n+1}\,dr
\int_0^{2\pi}K\!\left(s_\epsilon(r,\theta)\right)e^{+in\theta}\,d\theta.
\label{eq:Qn_epsilon_integral}
\end{equation}
The argument of the profile can then be written as
\begin{align}
s_\epsilon(r,\theta)={}&(1-\epsilon\eta_0)
\bigl(r\cos\theta-\epsilon r_0\cos\theta_0\bigr)^2\nonumber\\
&+(1+\epsilon\eta_0)
\bigl(r\sin\theta-\epsilon r_0\sin\theta_0\bigr)^2.
\label{eq:elliptical_argument}
\end{align}
The function $K$ is the circular projected density profile; its derivatives below are taken with respect to its argument. The parameter $\epsilon$ is only a bookkeeping parameter; the physical ellipticity and displacement are specified by Eq.~(\ref{eq:epsilon_parameters}).

We expand the integral itself,
\begin{equation}
Q_n(\epsilon)=Q_n(0)+\epsilon Q_n'(0)
+\frac{\epsilon^2}{2}Q_n''(0)+O(\epsilon^3).
\label{eq:Qn_taylor}
\end{equation}
For $n\geq1$, $Q_n(0)=0$ because the unperturbed profile is circular. For a profile whose derivatives are regular at the origin, the usual differentiation under the integral conditions apply directly. The projected NFW profile requires a little more care: it has a logarithmic central cusp, so its derivatives are not bounded at $r=0$. We therefore understand the Taylor coefficient integrals below as improper integrals. One may first perform the expansion for $r\geq\delta>0$ and then take the limit $\delta\to0$. For the NFW and Einasto profiles considered here, the resulting weighted radial integrals through the required order converge at the lower endpoint. Thus no artificial core is required to make the multipole coefficients finite. In this improper integral sense, the coefficients are
\begin{equation}
Q_n^{(k)}(0)=\frac{1}{2\pi n}\int_0^1r^{n+1}\,dr
\int_0^{2\pi}e^{+in\theta}
\left.\frac{\partial^k}{\partial\epsilon^k}
K\!\left(s_\epsilon(r,\theta)\right)\right|_{\epsilon=0}
\,d\theta,
\label{eq:Qn_derivative_integral}
\end{equation}
where the lower endpoint is understood as the limit $\delta\to0$. The Taylor coefficients are consequently evaluated by differentiating the integrand away from the cusp and then taking the convergent improper integral limit.
To evaluate these derivatives, write
\begin{equation}
s_\epsilon=r^2+\epsilon A(r,\theta)
+\epsilon^2B(r,\theta)+O(\epsilon^3),
\label{eq:s_epsilon_expansion}
\end{equation}
where
\begin{align}
A(r,\theta)&=-2rr_0\cos(\theta-\theta_0)
-\eta_0 r^2\cos2\theta,\label{eq:A}\\
B(r,\theta)&=r_0^2+2\eta_0 rr_0\cos(\theta+\theta_0).
\label{eq:B}
\end{align}
The chain rule gives
\begin{align}
\left.\frac{\partial}{\partial\epsilon}K(s_\epsilon)\right|_{\epsilon=0}
&=A K'(r^2),\\
\left.\frac{1}{2}\frac{\partial^2}{\partial\epsilon^2}
K(s_\epsilon)\right|_{\epsilon=0}
&=B K'(r^2)+\frac12A^2K''(r^2).
\end{align}
Therefore the density expansion entering Eq.~(\ref{eq:Qn_epsilon_integral}) is
\begin{equation}
\Sigma_\epsilon
=K(r^2)+\epsilon A K'(r^2)
+\epsilon^2\left[B K'(r^2)
+\frac12 A^2K''(r^2)\right]+O(\epsilon^3).
\label{eq:sigma_taylor}
\end{equation}

For the multipole calculation, it is important to retain the full angular dependence before carrying out the angular integral. At first order,
\begin{equation}
\Sigma^{(1)}(r,\theta)
=-2rr_0 K'\left(\cos\theta\cos\theta_0
+\sin\theta\sin\theta_0\right)
-\eta r^2 K'\cos2\theta.
\label{eq:sigma_first_harmonics}
\end{equation}
At second order, the complete harmonic decomposition is
\begin{align}
\Sigma^{(2)}(r,\theta)={}&D_0(r)
+D_1(r)\left(\cos\theta\cos\theta_0
-\sin\theta\sin\theta_0\right)\nonumber\\
&+D_2(r)\left(\cos2\theta\cos2\theta_0
+\sin2\theta\sin2\theta_0\right)\nonumber\\
&+D_3(r)\left(\cos3\theta\cos\theta_0
+\sin3\theta\sin\theta_0\right)
+D_4(r)\cos4\theta,
\label{eq:sigma_second_harmonics}
\end{align}
where
\begin{align}
D_0(r)={}&r_0^2K'
+\left(r^2r_0^2+\frac{\eta^2r^4}{4}\right)K'',\label{eq:D0}\\
D_1(r)={}&\eta r_0\left(2rK'+r^3K''\right),\label{eq:D1}\\
D_2(r)={}&r^2r_0^2 K'',\label{eq:D2}\\
D_3(r)={}&\eta r^3r_0 K'',\label{eq:D3}\\
D_4(r)={}&\frac{\eta^2r^4}{4} K''.
\label{eq:D4}
\end{align}
The first order term contains the dipole generated by the displacement and the quadrupole generated by ellipticity. At second order, the displacement produces a correction to the quadrupole, the coupling between displacement and ellipticity generates the third harmonic, and the square of the ellipticity generates the fourth harmonic.

For the model in Eq.~(\ref{eq:sigma}), the inner potential multipoles are obtained by evaluating Eq.~(\ref{eq:Qn_definition}) with $\Sigma=\Sigma_\epsilon$. The normalization factor $1/(2\pi n)$ is essential because $Q_n$ denotes the coefficient of the potential multipole expansion, not the un-normalized Fourier weighted density integral. The angular selection follows, for example, from
\begin{equation}
\int_0^{2\pi}\cos(n\theta-\alpha)e^{+in\theta}\,d\theta
=\pi e^{+i\alpha}.
\label{eq:harmonic_integral}
\end{equation}
Equivalently, the cosine and sine terms each contribute through their complex Fourier coefficient, with the angular integration giving $2\pi$ times that coefficient.

The leading terms in the four lowest moments are most transparently written in terms of the physical displacement coordinates and ellipticity defined in Eq.~(\ref{eq:epsilon_parameters}):
\begin{align}
Q_1={}&-(x_c+i y_c)
\int_0^1r^3 K'(r^2)\,dr+O(\epsilon^2),\label{eq:Q1}\\
Q_2={}&-\frac{\eta}{4}
\int_0^1r^5 K'(r^2)\,dr+O(\epsilon^2),\label{eq:Q2}\\
Q_3={}&\frac{\eta}{6}(x_c+i y_c)
\int_0^1r^7 K''(r^2)\,dr+O(\epsilon^3),\label{eq:Q3}\\
Q_4={}&\frac{\eta^{\,2}}{32}
\int_0^1r^9 K''(r^2)\,dr+O(\epsilon^3).\label{eq:Q4}
\end{align}
In Eqs.~(\ref{eq:Q1}) and (\ref{eq:Q4}), we have returned from the bookkeeping variables to the physical parameters $\eta$, $x_c$, and $y_c$.
For $Q_3$, the real and imaginary contributions arise respectively from the $\cos3\theta\cos\theta_0$ and $\sin3\theta\sin\theta_0$ terms in Eq.~(\ref{eq:sigma_second_harmonics}); together they give the phase of the physical displacement, $x_c+i y_c$. For $Q_4$, the leading term is proportional to $\cos4\theta$ and its sine projection vanishes in the coordinate system where the ellipse is aligned with the axes. The second order off centering correction to $Q_2$ is
\begin{equation}
Q_2^{(2)}=\frac{(x_c+i y_c)^2}{4}
\int_0^1r^5 K''(r^2)\,dr,
\end{equation}
but it does not alter the leading order hierarchy in Eqs.~(\ref{eq:Q1}) and (\ref{eq:Q4}).

For compactness, define the radial integrals
\begin{align}
I_1&=\int_0^1r^3 K'(r^2)\,dr,
&I_2&=\int_0^1r^5 K'(r^2)\,dr,\label{eq:I12}\\
J_3&=\int_0^1r^7 K''(r^2)\,dr,
&J_4&=\int_0^1r^9 K''(r^2)\,dr.\label{eq:J34}
\end{align}
The leading ratios are then
\begin{align}
\left|\frac{Q_3}{Q_1}\right|
&=\eta\,\mathcal C_{31},
&\mathcal C_{31}&=\frac16\left|\frac{J_3}{I_1}\right|,
\label{eq:C31}\\
\left|\frac{Q_4}{Q_2}\right|
&=\eta\,\mathcal C_{42},
&\mathcal C_{42}&=\frac18\left|\frac{J_4}{I_2}\right|.
\label{eq:C42}
\end{align}
The displacement parameter $r_0$ cancels from the first ratio, provided the displacement is non zero so that the leading dipole is defined. The second ratio is independent of $r_0$ at this order. Both ratios are proportional to the physical ellipticity $\eta=\epsilon\eta_0$; the bookkeeping parameter therefore appears in the final ratios only through the physical ellipticity. The factors $1/6$ and $1/8$ include the multipole order normalization $1/n$ in the potential coefficients defined in Eq.~(\ref{eq:Qn_definition}).

The radial integrals can be expressed in terms of ordinary moments of the circular projected profile. Let $u=r^2$ and define
\begin{equation}
M_j=\int_0^1u^j K(u)\,du,
\end{equation}
where $K(u)$ denotes the unperturbed circular surface density as a function of $u$. Integration by parts gives
\begin{align}
I_1&=\frac12\left[K(1)-M_0\right],\label{eq:I1_mom}\\
I_2&=\frac12\left[K(1)-2M_1\right],\label{eq:I2_mom}\\
J_3&=\frac12\left[K'(1)-3K(1)+6M_1\right],\label{eq:J3_mom}\\
J_4&=\frac12\left[K'(1)-4K(1)+12M_2\right].\label{eq:J4_mom}
\end{align}
These expressions make explicit that the hierarchy depends on the radial structure of the projected halo, including the profile values and slope at the Einstein radius boundary. The $Q_n$ used here are the normalized potential coefficients; the corresponding un-normalized density Fourier moments are $2\pi n Q_n$. For profiles with a central cusp, the original integrals in Eqs.~(\ref{eq:I12}) and (\ref{eq:J34}) are understood as convergent improper integrals; the weighted integrands are integrable for the NFW and Einasto cases considered below.

\subsection{NFW and Einasto estimates}
\label{sec:profile_estimates}

We estimate the coefficients in Eqs.~(\ref{eq:C31}) and (\ref{eq:C42}) for two commonly used spherical halo profiles. The overall surface density normalization cancels from both coefficients. We therefore write the circular projected profile in the form
\begin{equation}
K(u)=\mathcal S(x_E\sqrt{u}),
\qquad x_E=\frac{R_E}{r_s},
\label{eq:F_profile}
\end{equation}
where $r_s$ is the NFW scale radius or, for the Einasto model, the radius $r_{-2}$ at which the logarithmic three-dimensional density slope is $-2$.

For NFW, the dimensionless projected profile is $\Sigma_{\rm NFW}(R)=2\rho_s r_s f_{\rm NFW}(x)$, with $x=R/r_s$ and \citep{WrightBrainerd2000}
\begin{equation}
f_{\rm NFW}(x)=
\begin{cases}
\displaystyle\frac{1-\frac{2}{\sqrt{1-x^2}}\operatorname{artanh}\sqrt{\frac{1-x}{1+x}}}{x^2-1},&x<1,\\[1.2ex]
\displaystyle\frac13,&x=1,\\[1.2ex]
\displaystyle\frac{1-\frac{2}{\sqrt{x^2-1}}\arctan\sqrt{\frac{x-1}{1+x}}}{x^2-1},&x>1.
\end{cases}
\label{eq:nfw_projected}
\end{equation}
For an Einasto halo we use
\begin{equation}
\rho_{\rm E}(r)=\rho_{-2}
\exp\left\{-\frac{2}{\alpha}
\left[\left(\frac{r}{r_{-2}}\right)^\alpha-1\right]\right\},
\label{eq:einasto_3d}
\end{equation}
whose projected profile is obtained by line of sight integration,
\begin{equation}
\Sigma_{\rm E}(R)=2\rho_{-2}r_{-2}
\int_0^\infty
\exp\left\{-\frac{2}{\alpha}
\left[(x^2+t^2)^{\alpha/2}-1\right]\right\}dt,
\qquad x=\frac{R}{r_{-2}}.
\label{eq:einasto_projected}
\end{equation}
The projected Einasto profile can be evaluated by direct line of sight integration or by the analytic representations discussed by \citet{RetanaMontenegro2012}. The Einasto shape parameter depends on halo mass and formation history. Values around $\alpha\simeq0.16$ are representative of galaxy scale halos, while values approaching $0.3$ occur for more massive halos \citep{Gao2008}. We therefore use $\alpha=0.16$ and $0.30$ as representative cases, rather than treating either value as universal.

We evaluated the one dimensional radial moments in Eqs.~(\ref{eq:I1_mom}) and (\ref{eq:J4_mom}) using the projected profiles above. Table~\ref{tab:profile_coefficients} lists the coefficients $\mathcal C_{31}$ and $\mathcal C_{42}$ for representative values of $x_E$. The range $0.05\leq x_E\leq1$ is intended as a broad sensitivity interval for the position of the Einstein radius relative to the halo scale radius; it is not asserted to be the measured distribution of $x_E$ in a particular lens sample.

\begin{table}[ht]
\centering
\caption{Profile dependent coefficients in the leading ratios of the inner potential multipoles, $|Q_3/Q_1|=\eta\mathcal C_{31}$ and $|Q_4/Q_2|=\eta\mathcal C_{42}$, where $Q_n=a_n+ib_n$ follows the normalization of Alard (2008, 2010), and $\eta$ is the physical ellipticity parameter in the adopted quadratic form parametrization.}
\label{tab:profile_coefficients}
\begin{tabular}{llccc}
\toprule
Profile & $\alpha$ & $x_E=0.05$ & $x_E=0.2$ & $x_E=1$ \\
\midrule
NFW & -- & $0.084,\ 0.084$ & $0.087,\ 0.089$ & $0.098,\ 0.109$ \\
Einasto & $0.16$ & $0.087,\ 0.087$ & $0.092,\ 0.097$ & $0.095,\ 0.110$ \\
Einasto & $0.30$ & $0.073,\ 0.068$ & $0.085,\ 0.084$ & $0.099,\ 0.108$ \\
\bottomrule
\end{tabular}
\begin{flushleft}
\footnotesize Each table entry gives $(\mathcal C_{31},\mathcal C_{42})$.
\end{flushleft}
\end{table}

Across the grid $0.05\leq x_E\leq1$ and $0.16\leq\alpha\leq0.30$, the coefficients remain within the approximate ranges
\begin{equation}
0.073\lesssim\mathcal C_{31}\lesssim0.099,
\qquad
0.068\lesssim\mathcal C_{42}\lesssim0.110.
\label{eq:coeff_ranges}
\end{equation}
These ranges are relatively insensitive to the overall density normalization and vary smoothly with the position of the Einstein radius within the scale radius. We consider the interval $0.2\leq\eta\leq0.4$ as a plausible illustrative range for projected halo flattening. This is not intended as a universal interval: the observed stellar shapes in strong lens samples, including the SL2S sample used for the acceleration scale estimates, and the shape distributions of dark matter and baryonic components in hydrodynamical simulations provide complementary empirical and theoretical context \citep{Sonnenfeld2013a,Velliscig2015}. In the adopted quadratic form parametrization, this interval corresponds to projected axis ratios $q=\sqrt{(1-\eta)/(1+\eta)}\simeq0.65$ to $0.82$. Over this range, the smooth halo model predicts approximately
\begin{equation}
0.015\lesssim\left|\frac{Q_3}{Q_1}\right|\lesssim0.040,
\qquad
0.014\lesssim\left|\frac{Q_4}{Q_2}\right|\lesssim0.044,
\label{eq:moment_ratio_estimates}
\end{equation}
These estimates quantify the expected hierarchy for the simple smooth elliptical, off centered model. They are not universal CDM predictions: real halos can have radially varying shapes, triaxiality, isophotal deviations, interactions, and substructure. The calculation provides a baseline against which such additional contributions can be assessed.

\subsection{Substructure and higher-order power}
\label{sec:substructure}

The smooth halo calculation indicates that global ellipticity and a modest center displacement generate relatively small ratios of higher order to lower order potential multipole moments. A larger high order signal may therefore point to departures from the smooth elliptical description. Localized substructures are particularly relevant. In the perturbative formalism, a substructure close to the critical circle can produce substantial high frequency power in the angular fields. \citet{Alard2008} showed that the power spectra of the two perturbative fields have a characteristic relation for localized substructure perturbations, distinct from the relation expected for self-similar higher order distortions. The detailed analysis of that signature is given in the earlier work; here it is sufficient to note that the higher order inner multipoles offer a natural observational channel through which such perturbations may be investigated.

The comparison must nevertheless distinguish substructure from other departures from a global ellipse. Radially varying isophote shapes, boxy or disky distortions, nearby companions projected within the Einstein aperture, and line of sight structures can also contribute to higher harmonics. The smooth NFW/Einasto estimates should therefore be regarded as a reference expectation, not as a unique discriminator. The combination of the full harmonic content of both perturbative fields, their inner/outer separation, and population statistics will be needed to interpret an observed excess of higher-order power.

\section{Discussion}
\label{sec:discussion}

The framework developed here is intended to test, rather than assume, a non radial extension of the RAR. Its central advantage is the combination of a direct reconstruction of angular fields with a separation of the projected mass inside and outside the Einstein circle. The fields $f_0(\theta)$ and $f_1(\theta)$ are connected to the observed image morphology and do not require a parametric mass model. The mass sheet degeneracy acts as a common re-scaling of these fields; normalization removes that amplitude, while a separate normalization of the baryonic field removes a constant stellar mass to light factor.

The inner multipoles are the primary quantities for the RAR comparison because they describe the projected mass within the Einstein aperture. The outer terms describe the environment and can be affected by external shear and by the selection of strongly lensed systems. This distinction must be preserved both in individual reconstructions and in population statistics. The projected nature of the observables also limits the interpretation: the diagrams compare reduced two dimensional fields, not local three dimensional accelerations.

The inner quadrupole orientation provides a test of the relative geometry of the baryonic and total projected distributions. Higher order moments provide a complementary diagnostic. The smooth elliptical, off centered halo calculation gives a baseline whose amplitude depends on the radial profile and on the Einstein radius relative to the halo scale radius. For the NFW and Einasto profiles considered here, the predicted ratios are small for moderate ellipticity. An excess of higher order power may therefore indicate substructure or other departures from a smooth ellipse, although the detailed harmonic pattern is needed to distinguish these possibilities. The effects of localized substructures on the perturbative fields have been discussed by \citet{Alard2008}.


The projected relations introduced here are not intended to reconstruct uniquely a three dimensional non radial acceleration relation. Their purpose is different. They provide an observationally accessible test of whether such a relation leaves a statistical imprint in the projected internal gravitational field. A local three dimensional relation need not be preserved point by point under line of sight projection, which may dilute or partly cancel its angular signatures. Nevertheless, if a proposed three dimensional extension entails a strong and generic correspondence between the baryonic and total non radial acceleration fields over the relevant range of radii and orientations, one would expect this correspondence to leave a statistical imprint in their projected counterparts. Consequently, the absence of a detectable correspondence in the projected inner fields, once projection effects and observational uncertainties are accounted for, would place strong constraints on—and potentially rule out—classes of three dimensional non radial extensions that predict such a projected signal. In this sense, the framework developed here should be viewed primarily as a falsification test of non radial RAR hypotheses rather than as a reconstruction of a unique three dimensional relation. It is also worth emphasizing that projected correlations between baryonic and total mass distributions are not specific to a particular theory of gravity. In the standard CDM framework, baryons and dark matter evolve within the same gravitational potential and are therefore expected to exhibit some statistical correspondence in their projected structures, even if their detailed distributions and orientations differ. Similar correlations are also expected in modified gravity theories.

\subsection{Extension to group scale lenses}
\label{sec:groups}

The method also applies to group scale and low mass cluster scale deflectors. The SL2S survey contains such systems; in particular, the SARCS sample was designed to identify lenses on mass scales intermediate between individual galaxies and massive clusters, and includes numerous group scale candidates \citep{More2012}.

This mass regime is especially interesting because the tight RAR established for ordinary galaxies does not appear to extend straightforwardly to groups and clusters. Analyses of X-ray cluster samples report substantially larger scatter and possible changes in the form of the relation, although these results depend on assumptions about hydrostatic equilibrium and the baryonic mass inventory \citep{ChanDelPopolo2020,ChanLaw2022}. Moreover, MOND leaves a well known residual mass discrepancy in X-ray groups and clusters \citep{Angus2008}. Applying the present method to group scale lenses would allow us to test whether these departures from the galaxy scale radial phenomenology have a counterpart in the angular structure of the projected field. This could provide an independent clue to the nature and spatial distribution of the additional gravitating component. In some groups and clusters, the hot gas can contribute significantly to the baryonic mass distribution and may therefore be a source of uncertainty when the baryonic field is inferred from stellar light alone; this is a system dependent issue, not a limitation that applies to every lens. The inner outer separation is particularly important here, since the outer field can receive substantial contributions from the surrounding group or cluster environment.

\subsection{The statistical opportunity offered by Euclid}
\label{sec:euclid}

The population statistics required for this program are beyond the reach of current small, well characterized lens samples. Euclid offers the prospect of a substantial increase in both sample size and imaging uniformity. Its Q1 strong lensing catalogue contains 497 galaxy--galaxy lens candidates over 63 square degrees, including 250 grade-A candidates \citep{Euclid Q1}. The discovery analysis forecasts more than one hundred thousand high confidence candidates over the complete Wide Survey \citep{Euclid Q1}.

These forecast candidates should not be confused with systems immediately suitable for the present analysis. Only a subset will have sufficiently resolved images, reliable redshifts, and the photometric information needed to reconstruct the baryonic field. Complementary spectroscopy and a careful characterization of the lens selection function will also be required. The effective sample will therefore depend on image quality, data processing, source lens separation, and the availability of the ancillary information needed for each system. Euclid's main contribution will be the population scale and the possibility of studying a large, consistently selected sample over a very wide area.

The reconstruction accuracy discussed above provides an indication of the practical precision that can be reached on individual systems. In particular, the measurement of the relative baryonic total quadrupole orientation in SL2S02176-0513 by \citet{Alard2010}, with an uncertainty of about five degrees, demonstrates that this type of angular observable can be measured in high resolution strong lensing data. A rough extrapolation suggests that a comparable accuracy may be achievable for suitable Euclid lenses. This expectation should, however, be regarded as provisional: it has not yet been tested on Euclid observations, and the actual precision will depend on the image quality, the treatment of the image data, and the details of the reconstruction. We therefore do not attempt to give a quantitative forecast for the Euclid measurement errors at this stage.

The main opportunity offered by Euclid is consequently statistical rather than a presumed improvement in the precision of each individual reconstruction. Even if the accuracy per lens is comparable to that already demonstrated, a sufficiently large and homogeneous sample would make it possible to study distributions of quadrupole misalignment's, correlations between the normalized radial and tangential fields, and their dependence on characteristic acceleration, morphology, and environment. Population level analyses could also help distinguish generic trends from the scatter and system to system complexity present in individual lenses.

The combination is the key. The singular perturbative approach supplies a common, non model dependent reconstruction directly tied to image morphology; Euclid can potentially supply the population over which these consistently defined fields are compared. Together they offer the statistical leverage to investigate the radial and tangential angular relations across a broad range of characteristic accelerations, morphology's, and environments. The feasibility of the program is supported by the demonstrated precision of existing strong lensing measurements, while its full statistical realization will depend on the quality and suitability of the Euclid data products. This is what opens a new observational window on the RAR and its possible physical origin.

\section{Conclusions}
\label{sec:conclusions}

We have developed an observational framework for testing whether the RAR has a non radial counterpart. The method reconstructs two angular perturbative fields from strong lens image morphology, separates the projected mass inside and outside the Einstein circle, and compares normalized baryonic and total inner fields. The radial and tangential components define two complementary statistical diagrams. The multipole analysis adds information on the relative orientation of the baryonic and total distributions and on departures from smooth elliptical structure.

The approach does not assume that a non radial relation exists. Its purpose is to make the question observationally testable. A detected relation, a null result, or systematic dependence on acceleration, morphology, or environment would each constrain the physical meaning and domain of validity of the RAR. The extension to group scale lenses is particularly valuable: these systems probe a mass regime in which both the radial acceleration phenomenology and the residual mass discrepancy in MOND differ from the well studied galaxy case. Determining whether these differences are also reflected in the non radial projected field would provide an additional observational constraint on the origin of the RAR.

\begin{appendix}
\section{Convergence of the projected derivatives}
\label{app:convergence}

We establish the convergence of the line of sight integrals of the first and second derivatives of the Newtonian potential for a finite, sufficiently regular halo. The convergence of the integral of the potential itself is not required. Let the density have compact support inside $r'\leq R_0$ and finite total mass
\begin{equation}
M_h=\int\rho(\bm x')\,d^3x'.
\end{equation}
The potential is
\begin{equation}
\Phi(\bm x)=-G\int\frac{\rho(\bm x')}{|\bm x-\bm x'|}\,d^3x'.
\end{equation}
For a field point with $r=|\bm x|>R_0$,
\begin{equation}
|\bm x-\bm x'|\geq r-|\bm x'|\geq r-R_0.
\end{equation}
It follows that
\begin{equation}
|\partial_i\Phi(\bm x)|\leq\frac{GM_h}{(r-R_0)^2}.
\label{eq:first_derivative_bound}
\end{equation}
Using
\begin{equation}
\partial_i\partial_j\frac{1}{|\bm x-\bm x'|}
=\frac{3(x_i-x_i')(x_j-x_j')}{|\bm x-\bm x'|^5}
-\frac{\delta_{ij}}{|\bm x-\bm x'|^3},
\end{equation}
one obtains
\begin{equation}
|\partial_i\partial_j\Phi(\bm x)|
\leq\frac{4GM_h}{(r-R_0)^3}.
\label{eq:second_derivative_bound}
\end{equation}
Choose a fixed $r_1>R_0$ and define $c=1-R_0/r_1>0$. For $r\geq r_1$, $r-R_0\geq cr$, so
\begin{equation}
|\partial_i\Phi|\leq\frac{GM_h}{c^2r^2},
\qquad
|\partial_i\partial_j\Phi|\leq\frac{4GM_h}{c^3r^3}.
\end{equation}
Along a line of sight at fixed projected radius $R>0$, $r^2=R^2+z^2$. The part of the line of sight for which $r<r_1$ is a finite interval and gives a finite contribution for a sufficiently regular density distribution. On the two tails, the preceding bounds are integrable because
\begin{equation}
\int_{-\infty}^{+\infty}\frac{dz}{R^2+z^2}=\frac{\pi}{R},
\qquad
\int_{-\infty}^{+\infty}\frac{dz}{(R^2+z^2)^{3/2}}=\frac{2}{R^2}.
\end{equation}
The line of sight integrals of the first and second derivatives therefore converge absolutely. The compact-support assumption gives a simple explicit proof. The same convergence also holds for the untruncated NFW profile used in the illustrative calculation: its outer density behaves as $r^{-3}$, so its enclosed mass grows only logarithmically and its first and second potential derivatives decay as $\ln r/r^2$ and $\ln r/r^3$, respectively. Both tails are integrable along the line of sight. The Einasto profile decays still more rapidly.

Finally, integrating the three-dimensional Poisson equation gives
\begin{align}
\int_{-\infty}^{+\infty}(\Phi_{,xx}+\Phi_{,yy})\,dz
&=4\pi G\int_{-\infty}^{+\infty}\rho\,dz
-\left[\Phi_{,z}\right]_{-\infty}^{+\infty}\\
&=4\pi G\Sigma(x,y),
\end{align}
since $\Phi_{,z}\rightarrow0$ as $z\rightarrow\pm\infty$. This establishes Eq.~(\ref{eq:projected_poisson}) without requiring convergence of $\int\Phi\,dz$.

\end{appendix}

\end{document}